\documentclass{PoS}

\usepackage{transparent}
\usepackage{amsmath}
\usepackage{amssymb}
\usepackage{bm}
\usepackage[table,x11names]{xcolor}
\usepackage{graphicx}
\usepackage{gensymb}
\usepackage[font=footnotesize,labelfont=bf]{caption}
\usepackage{subcaption}

\title{Discovery of the TeV Emission from the Jet Interaction Regions of SS~433 with HAWC}

\ShortTitle{HAWC SS~433 Follow-Up}

\author{Chang Dong Rho\\
        University of Rochester\\
        E-mail: \email{no397@naver.com}}
        
\author{\speaker{Hao Zhou}\\
       Los Alamos National Laboratory\\
       E-mail: \email{hzhou1@mtu.edu}}
       
\author{Segev BenZvi\\
        University of Rochester\\
        E-mail: \email{sybenzvi@pas.rochester.edu}}

\author{For the HAWC Collaboration\\
		For a complete author list, see https://www.hawc-observatory.org/collaboration/icrc2019.php.}

\abstract{The High Altitude Water Cherenkov (HAWC) observatory recently published the discovery of SS~433 as a TeV source, reporting the observation of multi-TeV gamma-ray emission from the jet interaction regions e1 and w1, suggesting in-situ particle acceleration. This showed the first direct evidence of acceleration in jets at energies greater than a few TeV. SS~433 was the first microquasar to be discovered and is still considered special in that the accretion is supercritical and the luminosity of the system is very high ($\sim10^{40}$~erg~s$^{-2}$). The lobes of the supernova remnant W~50 in which the jets terminate, about 40 parsecs from the central binary, are expected to accelerate charged particles, and indeed radio and X-ray emission consistent with electron synchrotron emission in a magnetic field have been observed. SS 433 has also been a strong candidate for hadronic acceleration due to spectroscopic evidence of ionized nuclei in the inner jets. However, multiwavelength fits including the HAWC measurements favor the leptonic production of the observed gamma rays. Here, we present new follow-up measurements of the jet interaction regions of SS~433 using the most recent data from HAWC. 
}

\FullConference{36th International Cosmic Ray Conference -ICRC2019-\\
		July 24th - August 1st, 2019\\
		Madison, WI, U.S.A.}

\begin{document}

\section{Introduction}\label{sec:intro}
Active galactic nuclei (AGN) produce powerful jets of ionized matter which are believed to accelerate the highest energy hadronic cosmic rays. However, direct observation and tracking of cosmic rays from AGN is not yet achievable as they are electrically charged. Hence, observing gamma rays from jets would provide alternative evidence of such particle acceleration. Unfortunately, most AGN are too far away for us to spatially resolve their jets with the current generation of gamma-ray telescopes. However, it is possible to observe jets in microquasars, extreme star systems in the Milky Way that behave like AGN in miniature.

SS~433 is a well-known X-ray binary, discovered in 1979. It is famous for being the first astrophysical object to be classified as a microquasar. The binary is located at ($\alpha=19^{\text{h}}$:11$^{\text{m}}$:49.6$^{\text{s}}$, $\delta={+}$04$^{\circ}$58'57.6''), $\sim5.5$~kpc from Earth. SS~433 consists of a supergiant star that overflows its Roche lobe, accreting material onto the compact object (either a black hole or neutron star) in orbit with the supergiant~\cite{Fabrika:2006casp}. SS~433 is considered to be peculiar because the whole system is ``cocooned'' in a supernova remnant, W~50 and it has relativistic jets beamed perpendicular to the accretion disk of its compact object with a bulk velocity of $0.26c$, that interact with the remnant~\cite{Fabrika:2006casp}. As the jets of energetic particles terminate within W~50, these jet interaction / termination regions (approximately 40~pc away from the binary) can provide an ideal environment for energetic particle acceleration enough to produce very-high-energy (VHE) gamma rays.

For SS~433, the jet interaction regions are visible in X-rays at the jet termination regions, $\sim40$~pc away from the central binary. Previous study of the jet interaction regions (e1, e2, e3, w1, and w2) in X-rays, monitored by R\"{o}ntgensatellit (ROSAT) and the Advanced Satellite for Cosmology and Astrophysics (ASCA)~\cite{Safi-Harb:1997apj}, show that excess X-ray emission is being produced far away from the central engine, near the jet termination regions. Another interesting observational phenomena to be noted about SS~433 is its iron emission lines at 6.8~keV~\cite{Margon:1984araa}. The presence of heavier nuclei indicates that hadronic acceleration is possibly taking place in the jets of SS~433.

In 2018, the SS~433 jet lobes at the known X-ray jet termination regions were successfully observed by HAWC in the VHE gamma-ray regime, which marks the first and the only direct evidence of microquasar jet emission measured in VHE gamma rays \cite{ss433paper}. The multi-source analysis with the original 33~months of HAWC data prefers the leptonic origin of the gamma-ray photons, though the hadronic-only model is not completely ruled out. Furthermore, the HAWC observation supports the acceleration site to be near the jet interaction regions, well away from the central binary, without a clear understanding of the particle acceleration mechanism. Here, we show follow-up HAWC measurements of SS~433 jet termination regions.

\section{HAWC}\label{sec:hawc}
The High Altitude Water Cherenkov (HAWC) Observatory is a ground array located at latitude of 19\degree N and at an altitude of 4,100 meters in Sierra Negra, Mexico. HAWC consists of 300 water Cherenkov detectors (WCDs) covering a large geometrical area of 22,000 m$^2$. Of the 300 deployed tanks, 294 have been instrumented \cite{hawccrabpaper}. Each WCD has a light-tight polypropylene bladder filled with 200,000 litres of purified water. The bladder is encased in a steel tank. At the bottom of each WCD there are three 8-inch Hamamatsu R5912 photomultiplier tubes (PMTs) oriented in an equilateral triangle and one 10-inch R7081-HQE PMT anchored at the center.

By combining the location and the time of each PMT triggered by an air shower, the core position and the angle at which the primary particle has generated the air shower is reconstructed to locate and identify the primary particle type. Simple topological cuts are applied to discriminate the air showers produced by hadronic cosmic rays from $\gamma$-ray air showers. For a detailed explanation of the event reconstruction, see \cite{smi15}.

The air shower trigger rate of HAWC is approximately 25kHz, more than 99.9\% originating from cosmic rays. At HAWC's altitude, a vertical shower from a 1 TeV photon will have about 7\% of the original photon energy when the shower reaches the tanks. This ratio increases to around 28\% at 100 TeV \cite{hawccrabpaper}. The main source of background to $\gamma$-ray observation is the hadronic cosmic-rays. Therefore, individual $\gamma$-ray-induced air showers are distinguished from cosmic-ray showers using their topology. HAWC has a duty cycle $>95\%$ and a wide, unbiased field of view of ~2 sr. As such, HAWC is well-suited to study long-duration light curves of astronomical objects, making it an excellent detector to search for $\gamma$-ray binary sources \cite{seg15}. Also, the wide field of view allows us to carry out effective multi-source analyses to help with a number of serendipitous discoveries of sources in complex source-confused regions. Furthermore, being the highest energy gamma-ray observatory in operation currently, the discovery of SS 433 jet lobes is likely to be associated with the sensitivity of HAWC to the highest energy gamma-ray emission.

The previous energy determination technique uses the fraction of the hit PMTs per gamma-ray event to obtain the estimated energy of the original gamma-ray photon. However, this method is not correlated with energy enough since it does not take into account important parameters such as the zenith angle and how well-contained the shower is within the array. Also, for events above $\sim10$~TeV, almost the entirety of the PMTs is hit during an air shower, which makes computation of the dynamic range above 10~TeV very difficult. Therefore, HAWC has developed two new techniques that are more resilient to these issues, one of which is used for the follow-up analysis presented in this proceeding. For further information on the new energy estimators, please refer to \cite{eecrab}.

\section{Analysis Method}
The basic analysis procedure used to obtain the results presented in this work is essentially same as \cite{ss433paper}. Hence, for more detailed information on the analysis and on the previous results, please refer to \cite{ss433paper}.

The study of the multi-TeV gamma-ray emission from the SS~433 jet interaction regions has been carried out using maximum likelihood fitting~\cite{liffpaper,3mlpaper}. Specifically, because SS~433 is located so close to MGRO~J1908${+}$06 (a bright and extended TeV source), we carry out simultaneous fitting of the east lobe (at e1), west lobe (at w1), and MGRO~J1908${+}$06 (at $\alpha=19^{\text{h}}$:08$^{\text{m}}$:12$^{\text{s}}$, $\delta=06^{\circ}23'24''$). We have used a point source model for each lobe and the electron diffusion morphology~\cite{gemingapaper} for MGRO~J1908${+}$06. For all of the fitted sources, we have assumed a simple power-law spectral model:
\begin{equation}
    f=A\left(\frac{E}{E_{\text{piv}}}\right)^{-\alpha},
\end{equation}
where $f$ is flux, $A$ is flux normalization, $E$ is energy, $E_{\text{piv}}$ is pivot energy, and $\alpha$ is spectral index. For the analysis of the SS~433 jet lobes, index is fixed at 2.0 and the pivot energy at 20~TeV.

Furthermore, we have applied a special semi-circular region of interest (brighter region in Fig.~\ref{fig:ss433map}) to exclude the upper half of MGRO~J1908${+}$06 that is closer to the Galactic plane. Using this region of interest allows us to proceed effectively with the likelihood fitting without having to model the ambient Galactic diffuse emission and include it in the simultaneous likelihood fit.

The main difference between the follow-up measurements presented in this work and our previous work~\cite{ss433paper} is the dataset used. This work adopts the new energy estimator that contains 1,039~days of HAWC data, which only contains ``on-array'' events (i.e. gamma-ray events with their projected core landing on the main array of HAWC), whereas our previous analysis used 1,017~days of HAWC data that contains both sets of on-array and off-array events.

\section{Results}
Table~\ref{table:table_ss433} provides the fit results of the flux normalizations at the jet interaction regions e1 and w1. The first set is from \cite{ss433paper} with 1,017~days of HAWC data. The second set is the follow-up measurements using a new energy estimation technique. Although it has slightly more transit days in the data, because it only contains on-array events, the new dataset actually contains much less number of gamma-ray events. Note that all of the results are obtained from simultaneously fitting MGRO~J1908${+}$06 (extended electron diffusion morphology), east lobe (point source at e1), and west lobe (point source at w1).

\begin{figure}[!htb]
  \centering
  \includegraphics[width=1.0\textwidth]{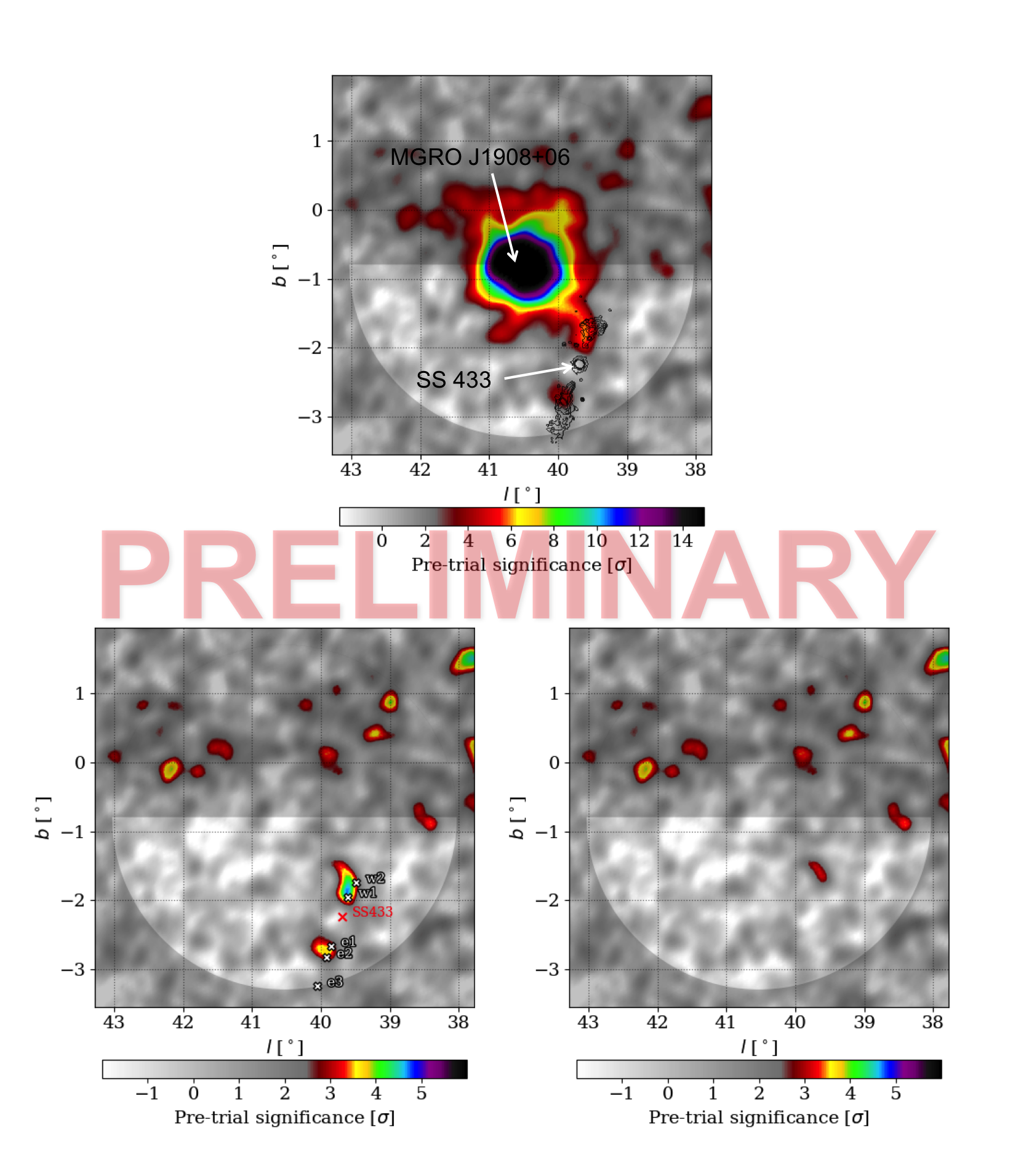}
  \caption{{\bf Top:} The original significance map of the SS 433 region. {\bf Bottom Left:} The residual map after subtracting the fitted MGRO J1908+06 model. {\bf Bottom Right:} The residual map after subtracting the fitted MGRO J1908+06 and the SS 433 lobes model. Although the plots here show successive subtraction of source models, the joint likelihood fit is actually performed simultaneously. Plots are only presented this way to better illustrate the residual emission.
  }
  \label{fig:ss433map}
\end{figure}

Fig.~\ref{fig:ss433map} presents the HAWC significance maps of the SS~433 region. The top plot shows the significance map without any source subtraction. The large bright blob of emission in the center is the observed emission from MGRO~J1908${+}$06. To the ``south-east'' of MGRO~J1908${+}$06 is the west lobe of SS~433. Then there is a gap of significant excess where the central binary is known to be. Further down is another hotspot corresponding to the east lobe. The HAWC hotspots are spatially in coincidence with the X-ray contours detected by ROSAT \cite{rosat} as shown in the plot. The bottom left plot is the residual map produced by simultaneously fitting the lobes and MGRO~J1908${+}$06 but subtracting only MGRO~J1908${+}$06 to help us visualize the fitted lobes. The known positions of the X-ray interaction regions as well as the central binary have also been labeled on the plot. The final bottom right plot shows the residual map after all the fitted sources have been subtracted. The region of interest now appears to be empty with no high significant excess nor any signs of over-subtraction.

Note that the new follow-up flux measurements that use the HAWC energy estimator are missing the systematic uncertainties in Table~\ref{table:table_ss433} since the study is still ongoing.

\bigskip
\begin{table}[htb!]
\centering
  {\bf Fits to the TeV emission from SS~433 using nested point source models}
  \begin{tabular}{cp{0.15\textwidth}p{0.25\textwidth}p{0.25\textwidth}p{0.15\textwidth}}
    \hline
    & Lobe & Position\newline (RA, Dec)
       & $dN/dE$ at 20 TeV\newline [$10^{-16}$ TeV$^{-1}$cm$^{-2}$s$^{-1}$] & TS 
    \\
    \hline
    \multicolumn{5}{l}{\bf Fractional hit bin results (published results).}
    \\
    & e1 & 19:13:37\newline 04$^\circ$55'48'' & $2.4^{+0.6+1.3}_{-0.5-1.3}$ & 41.2
    \\
    & w1 & 19:10:37 \newline 05$^\circ$02'13'' & $2.1^{+0.6+1.2}_{-0.5-1.2}$ &
    \\
    \multicolumn{5}{l}{\bf Energy estimator (energy estimator; on-array only).}
    \\
    & e1 & 19:13:37\newline 04$^\circ$55'48'' & $2.5^{+1.1}_{-0.8}$ & 30.0
    \\
    & w1 & 19:10:37 \newline 05$^\circ$02'13'' & $3.5^{+1.2}_{-0.9}$ &
    \\
    \hline
  \end{tabular}
  \caption{Fits to the TeV emission from SS 433 jet lobes using nested point source models. Contains simultaneous fits with fixed positions. The first set of results is with fractional hit bins from \cite{ss433paper} and the second set is with an energy estimator. The first set of uncertainties in the third column is the statistical uncertainties, followed by the systematic uncertainties. The energy estimator results are currently missing the systematic uncertainties.}
  \label{table:table_ss433}
\end{table}

\begin{figure}[!htb]
  \centering
  \includegraphics[width=1.0\textwidth]{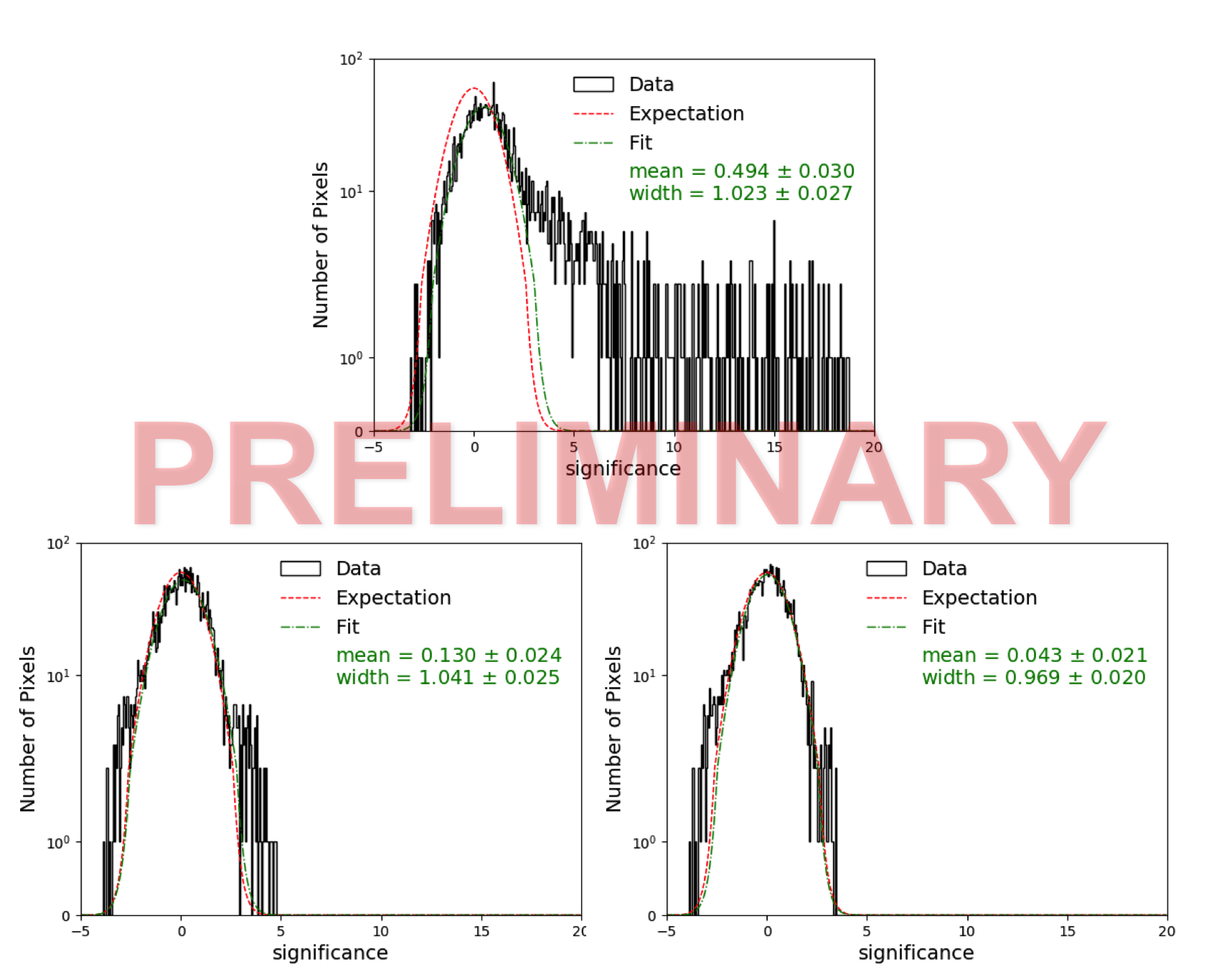}
  \caption{Distribution of pixel significance in the semi-circular RoI. The best fit Gaussian is in green dot-dashed curve. {\bf Top:} Corresponds to the pixels in the semi-circular RoI of Fig.~\ref{fig:ss433map} (top). No subtraction. {\bf Bottom Left:} Corresponds to the pixels in the semi-circular RoI of Fig.~\ref{fig:ss433map} (bottom left). Only MGRO~J1908+06 subtracted. {\bf Bottom Right:} Corresponds to the pixels in the semi-circular RoI of Fig.~\ref{fig:ss433map} (bottom right). Simultaneously fitted MGRO~J1908+06 and the SS~433 jet lobes subtracted, which the residual map is expected to be compatible with an empty background map (red dot-dashed curve).
  }
  \label{fig:ss433hist}
\end{figure}

\section{Discussion}
Fig.~\ref{fig:ss433hist} shows three significance distribution histograms corresponding to the semi-circular region of interest of the three significance maps in Fig.~\ref{fig:ss433map}. Each histogram has a green dotted curve that shows the best fit Gaussian to the distribution and a red dotted curve that shows the expected ``background-only'' distribution. In other words, when a set of significance values from all of the pixels in a region of interest is collected that contains no gamma-ray source, the red and green curves should match. This is the result we would expect when all the sources have been fitted and subtracted correctly from a region of interest. The first (top) histogram shows a distribution with a huge high significance tail since our region of interest contains emission from MGRO~J1908${+}$06 and the lobes. After subtracting fitted MGRO~J1908${+}$06, the significance histogram (bottom left) shows much improvement, but a high significance tail still exists because the lobes have not been subtracted yet. Finally, after subtracting the lobes as well as MGRO~J1908${+}$06, the bottom right significance histogram shows a distribution that is consistent with background. This study shows that our modeling of the SS~433 region and the fitting procedure have all been carried out properly. Also, there are other noticeable hotspots outside the semi-circular region of interest. However, these can be ignored since the region is close to the Galactic plane, hence these hotspots are possibly associated with unresolved point sources in the Galactic plane as well as being under the influence of the Galactic diffuse emission.

As Table~\ref{table:table_ss433} shows, the analysis carried out for the new set of data with the energy estimator is in good agreement with our previous results, which supports our published observation of the SS~433 jet interaction regions. Since the energy estimator currently uses only the on-array events, we will need more cumulative data to study the lobes in more detail.

\section{Conclusion}
This work has provided experimental results on the study of TeV gamma-ray emission from the jet interaction regions of SS~433 using 1,039~days of HAWC data. The multi-source analysis of the SS~433 jet interaction regions agrees with our previous study of TeV gamma-ray emission from microquasar jets. Assuming the significance of the VHE gamma-ray excess from SS 433 grows with additional data, more detailed spectral analysis of the gamma-ray jet emission will be possible. The new follow-up measurements also support the leptonic origin of the gamma-ray photons since HAWC still observes the SS 433 jet lobes as point sources. This observation disfavors a possible hadronic origin (pion decay) of the TeV gamma rays since protons would have spread to a few degrees before emitting gamma rays. However, the hadronic-only model is not completely ruled out. For more information on the analysis procedure and physics interpretation of our results, please refer to \cite{ss433paper}.

\end{document}